\documentclass[journal]{IEEEtran}

\ifCLASSINFOpdf
\else
   \usepackage[dvips]{graphicx}
\fi
\usepackage{url}

\usepackage{amsmath,graphicx}
\usepackage{booktabs}
\usepackage{amsfonts}
\usepackage[linesnumbered,ruled,vlined]{algorithm2e}
\usepackage{algpseudocode}
\usepackage{array}
\usepackage{bm}
\usepackage{textcomp}
\usepackage{stfloats}
\usepackage{url}
\usepackage{verbatim}
\usepackage{multirow}
\usepackage{cite}

\begin{document}

\title{Controlling your Attributes in Voice}

\author{Xuyuan Li, Zengqiang Shang, Li Wang, and Pengyuan Zhang, \IEEEmembership{Member, IEEE}
\thanks{Xuyuan Li and Pengyuan Zhang are with the Laboratory of Speech and Intelligent Information Processing, Institute of Acoustics, Chinese Academy
of Sciences, Beijing 100190, China, and also with the University of Chinese
Academy of Sciences, Beijing 100049, China (e-mail: lixuyuan@hccl.ioa.ac.cn;
zhangpengyuan@hccl.ioa.ac.cn).}
\thanks{Zengqiang Shang and Li Wang are with the Laboratory of Speech and Intelligent Information Processing, Institute of Acoustics, Chinese Academy
of Sciences, Beijing 100190, China (e-mail: shangzengqiang@hccl.ioa.ac.cn;
wangli@hccl.ioa.ac.cn).}
}

\markboth{Journal of \LaTeX\ Class Files, Vol. 14, No. 8, August 2015}
{Shell \MakeLowercase{\textit{et al.}}: Bare Demo of IEEEtran.cls for IEEE Journals}
\maketitle

\begin{abstract}
Attribute control in generative tasks aims to modify personal attributes, such as age and gender while preserving the identity information in the source sample. Although significant progress has been made in controlling facial attributes in image generation, similar approaches for speech generation remain largely unexplored. This letter proposes a novel method for controlling speaker attributes in speech without parallel data. Our approach consists of two main components: a GAN-based speaker representation variational autoencoder that extracts speaker identity and attributes from speaker vector, and a two-stage voice conversion model that captures the natural expression of speaker attributes in speech. Experimental results show that our proposed method not only achieves attribute control at the speaker representation level but also enables manipulation of the speaker age and gender at the speech level while preserving speech quality and speaker identity.

\end{abstract}


\IEEEpeerreviewmaketitle



\section{Introduction}

\IEEEPARstart{P}{ersonal} attributes such as age and gender convey critical information in human-computer interactions. The ability to control these attributes in generative models to produce more diverse samples is important for the advancing areas such as identity recognition, self-supervised learning, and privacy preservation \cite{yang2022heterogeneous,liu2022controllable,shirai2019privacy,singh2024synthetic}. While techniques for controlling personal attributes in generated facial images have become increasingly mature \cite{he2019attgan,HOU2022209,ning2024icgnet}, methods for manipulating these attributes in speech remain underexplored.

The universal impact of the age, gender, and other attributes of speakers on their voice has been widely established \cite{leeper1995speech,mortensen2006age,lee2013novel,rilliard2024evolution}. Meanwhile, the accuracy of recognizing these attributes has significantly improved \cite{wang2017learning,tursunov2021age,burkhardt2023speech}, benefiting from the development of neural network models. In the field of speech generation, Some works \cite{teixeira2024privacy, luu2022investigating,janbakhshi2022adversarial} have explored how to decouple and remove the personal attributes in speaker representations extracted by speaker recognition (SR) models. However, at the speech level, both recent speaker cloning models \cite{li2024sf,du2024cosyvoice,anastassiou2024seed} and speech privacy protection models \cite{miao2024benchmark,panariello2024speaker,yao2024distinctive} focus primarily on controlling the overall speaker. While there have been efforts to generate virtual speakers that reflect certain speaker attributes with facial images \cite{park2024synthe,lee2024fvtts} or text prompts \cite{hai2024dreamvoice,chen2024generating}, these approaches still struggle to control specific speaker attributes and speaker identity independently.

This letter introduces a novel approach to controlling speaker attributes in speech. Without the need for parallel data, our proposed method enables sharing the expression of speaker attributes in speech across different speakers. The process is structured in two key steps: \textbf{1) speaker attributes extracting; 2) conditional voice conversion.} In the first step, we design a GAN-based speaker representation variational autoencoder (SRVAE), which decomposes the speaker representations extracted by the speaker identification model into the attributes and identity components and generates a new speaker representation based on attributes labels and identity component. In the second step, we propose a two-stage voice conversion (TSVC) model, which refines the attribute-dependent average speech into speaker-specific speech based on extracted components extracted from generated speaker representation.

We evaluate our method on two speaker attributes: age and gender. Similar to other speech generation tasks with emotion and style control, controlling speaker attributes in speech involves a trade-off between attribute label consistency, speaker identity consistency, and speech quality. In Sec. \ref{ra}, we presented a series of subjective and objective experiments to analyze our approach across these three aspects. Our contributions can be summarized as follows:

1) We propose a novel approach that, to the best of our knowledge, is the first to achieve age and gender control at the speech level, while preserving speaker identity information.

2) We proposed the SRVAE, which can generate non-existent speaker representations with predefined age and gender labels, and extract attributes embedding from speaker representations.

3) We propose the TSVC, which enables the sharing of attribute expressions across different speakers in speech.


\section{Methods}

\begin{figure*}
\centerline{\includegraphics[width=\textwidth]{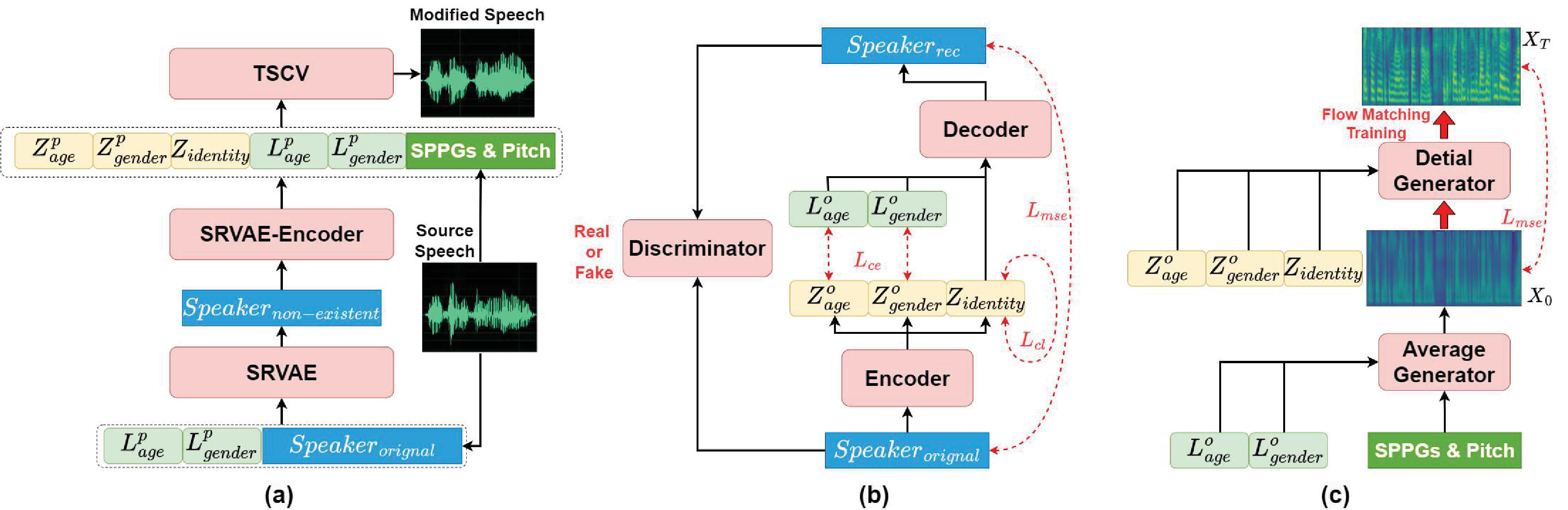}}
\caption{Overall workflow of our proposed method (a), detailed structure of SRVAE (b), and the detailed structure of TSVC (c). $L^p$ represents predefined attribute label, and $L^o$ represents original attribute label.}
\label{fig1}
\end{figure*}

\subsection{Overall Workflow}
\label{workflow}

Fig. \ref{fig1} (a) illustrates the overall workflow of our method. First, we apply the approach described in \cite{promonet} to extract sparse phonetic posteriorgrams (SPPGs) and pitch sequences from the source speech, which serve as the semantic features. Additionally, a WavLM-based SR model \cite{chen2022wavlm} is used to extract the original speaker vector from the source speech. Next, SRAVE generates a non-existent speaker vector based on the original speaker vector and predefined attribute labels. To capture more commonality and variation across different speakers, we use the encoder of SRAVE to decompose the synthetic speaker vector into age, gender, and identity embeddings as the conditions for TSVC, rather than directly using the speaker vector. Finally, the TSCV model conditions on these embeddings and the predefined attribute labels to transform the semantic features into speech with modified attributes.


\subsection{Speaker Representation Variational Autoencoder}
\label{srvae}

Fig. \ref{fig1} (b) illustrates the detailed architecture of SRAVE, a variational autoencoder trained with GAN. Specifically, we design an encoder consisting of three independent branches, which transforms the speaker vector into three latent embeddings: $\mathbf{z_{age}}$, $\mathbf{z_{gender}}$, and $\mathbf{z_{identity}}$. Each branch is composed of multiple residual blocks, each containing two linear layers with layer normalization and ReLU activation. After the age and gender branches, we employ a linear layer to output the probabilities of attribute labels for classification training. As for the identity branch, we introduce a contrastive loss to maximize the cosine similarity of identity embeddings between the same speaker and minimize the cosine similarity of identity embeddings between different speakers. On the decoder side, inspired by \cite{pan2021disentangled}, we use attribute labels, $\mathbf{l_{age}}$ and $\mathbf{l_{gender}}$, as input instead of attribute embeddings to improve training stability. After the decoder module, we introduce a discriminator for generative adversarial training. Both the decoder and discriminator share the same backbone structure as the encoder branches.

Unlike previous works \cite{teixeira2024privacy, luu2022investigating,janbakhshi2022adversarial}, we do not introduce operations, such as gradient reversal or mutual information minimization, to enforce independence between attribute and identity embeddings. On the contrary, we design a cyclic consistency training strategy to enable attribute sharing between speakers, as shown in Alg. \ref{a1}. In the consistency training step, the encoder is frozen to guide the non-existent speaker vector generated by the decoder as close as possible to the real speaker vector domain.

\SetKwInput{KwData}{Require}
\begin{algorithm}
\caption{Cyclic Consistency Training}
\label{a1}
\KwData{\small {Encoder $E$ with parameter $\theta$, decoder $D$ with \\ parameter $\phi$, optimizers $g_{\theta}$ and $g_{\phi}$ for parameters $\theta$ and $\phi$\\ respectively, input speaker vector $\mathbf{x}$, output reconstructed\\ speaker vectors $\mathbf{x_{rec}}$.}}
\For{\small {\textnormal{each training iteration}}}{
\small {
$\mathbf{z^o_{age}}, \mathbf{z^o_{gender}}, \mathbf{z_{identity}} = E(\mathbf{x})$ \\
$\mathbf{x^o_{rec}} = D(\mathbf{l^o_{age}}, \mathbf{l^o_{gender}}, \mathbf{z_{identity}})$ \\
\If{train discriminator} {
\small{ Train one step for the discriminator.}
}
Calculate classifier loss $\mathcal{L}_{ce}$, contrastive loss $\mathcal{L}_{cl}$, reconstructed loss $\mathcal{L}_{mse}$, and adversarial loss $\mathcal{L}_{adv}$: ~~~~~~ \begin{center}$\mathcal{L}_{(\theta,\phi)} = \mathcal{L}_{ce} + \mathcal{L}_{cl} + \mathcal{L}_{mse} + \mathcal{L}_{adv}$ \end{center} 
\nl Update parameter $(\theta, \phi) \leftarrow g_{(\theta, \phi)}\nabla_{(\theta, \phi)}\mathcal{L}_{(\theta,\phi)}$. \\
\If{train consistency} { \small{
\nl Random sample fake attribute labels $\mathbf{l^f_{age}}$ and $\mathbf{l^f_{gender}}$.
\nl Freeze encoder $E$. 
\nl $\mathbf{x^{non-existent}_{rec}} = D(\mathbf{l^f_{age}}, \mathbf{l^f_{gender}}, \mathbf{z_{identity}})$ 
\nl Calculate classifier loss $\mathcal{L}_{ce}$, contrastive loss $\mathcal{L}_{cl}$, and adversarial loss $\mathcal{L}_{adv}$: ~~~~~~ \begin{center}$\mathcal{L}_{\phi} = \mathcal{L}_{ce} + \mathcal{L}_{cl} + \mathcal{L}_{adv}$ \end{center}  
\nl Update parameter $\phi \leftarrow g_{\phi}\nabla_{\phi}\mathcal{L}_{\phi}$.
}
}
}
}
\end{algorithm}

\vspace{-10pt}
%

\subsection{Two-stage Voice Conversion}
\label{tscv}

We treat speech generation as a two-stage process: 1) generating average speech based on predefined ages and genders, and 2) refining the average speech into speech from a specific speaker based on the given attribute and identity embeddings. As illustrated in Figure \ref{fig1} (c), in the first stage, we design an average generator that uses attribute labels, SPPGs, and pitch sequences to produce average acoustic features. The network of this generator consists of a Bi-LSTM layer and multiple convolutional residual blocks with AdaIN \cite{huang2017arbitrary}. Notably, we do not use attribute embeddings in the average generator. Since no independence-guided operations are applied in SRAVE, the attribute embeddings are speaker-dependent. We believe these embeddings can help the second-stage module share the expression of attributes in the speech of similar speakers, but they are not suitable for generating speaker-independent average acoustic features.

In the second stage, we design a detail generator, an ordinary differential equation (ODE) model trained with flow matching \cite{liu2022flow}, to model the mapping between the average acoustic features $X_0$ and the speaker-specific acoustic features $X_T$. The backbone of this generator follows the transformer-based network from SF-Speech \cite{li2024sf}. During training, a random Gaussian noise $\xi$ is added to $X_0$ to obtain a continuous middle state distribution, expressed as ${X_t = (1-t)(X_{0}+\xi) + tX_T}$. The attribute embeddings, identity embeddings, and differential step $t$ are then repeated along the time axis and added to $X_t$, expressed as $X^{'}_t$. Finally, to improve pronunciation accuracy, we concatenate SPPGs, pitch, and $X^{'}_t$ along the hidden dimension, serving as the input to the transformer-based network to predict the direction of ODE, $dX_t$.




\section{Experiment}

We consider four questions in our experiments: 1) Can SRVAE correctly extract attribute and identity embedding, and generate corresponding speaker vectors with given attribute labels? 2) Can the speech generated by TSVC be consistent with the predefined attribute labels while maintaining naturalness? 3) Can the modified speech still be recognized as the original speaker's voice? 4) Are SRVAE and TSVC effective for this task? The audio samples are available on our demo page \footnote{https://lixuyuan102.github.io/control-your-attributes-in-speech/}.

\subsection{Dataset}

We conduct our experiments on a subset of Voxceleb2\cite{chung2018voxceleb2} with age labels from \cite{tawara2021age}. This subset contains about 5,000 speakers from 168K videos. We divided age into seven bands: less than 12, greater than 12 and less than 25, greater than 25 and less than 40, greater than 40 and less than 55, greater than 55 and less than 65, greater than 65 and less than 75, and greater than 75.

\subsection{Implementation Details}


We set up three models for comparison, one baseline method and two ablation methods.  \textbf{Method 1} is an ODE model with attribute labels and the original speaker vector as conditions. \textbf{Method 2} removes the average generator in the proposed TSVC model and employs an ODE model to map Gaussian noise to the target acoustic features directly. \textbf{Method 3} eliminates the SRVAE, instead using attribute labels and the original speaker vector as conditions to train the TSVC. 

We employ 3 residual blocks with 512 hidden dimensions to form the encoder and decoder of SRVAE, and 2 for its discriminator. For TSVC, we employ 3 convolutional residual blocks with 512 hidden dimensions for the average generator and 8 transformer layers with 1024 hidden dimensions for the detail generator. A HiFi-GAN \cite{kong2020hifi} vocoder was trained to convert this acoustic feature into speech waveform. During training, the pitch sequences are normalized sentence by sentence to remove speaker-specific information.

All models were trained on 2 Nvidia V100 GPUs. We employed an AdamW optimizer with a learning rate starting at 0.0001 and linearly decaying to train SRVAE for 800K steps, and another AdamW optimizer with a peak learning rate of 0.0001, linearly warmed up for 5000 steps and decayed in cosine annealing over the rest steps, to train TSVC for 300K steps.


\begin{table}[htbp]
  \begin{center}
  \caption{Evaluation Results of Attribute and Identity Embedding on Real and Generated Speaker Vectors.}
  \label{tab1}
  \begin{tabular}{c|c c c c} 
  \toprule [1pt]
   Speaker Vector & ACC-age $\uparrow$ & ACC-gender $\uparrow$ & CSIS $\uparrow$ & CSID $\downarrow$\\
  \midrule 
        Real &96.3\%&99.6\%& 0.861& 0.208\\
        Generated &98.0\%&99.9\%& 0.827& 0.194\\
  \bottomrule [1pt]
  \end{tabular}
  \end{center}
  \vspace{-12pt}
\end{table}

\section{Results and Analysis}
\label{ra}
In this section, we evaluate the effectiveness of the proposed method from both the speaker vector and speech levels.

\subsection{Evaluations of Generated Speaker Vector}
We assessed the extracted attribute embeddings by measuring their classification accuracy on the classifier in the SRVAE encoder. As for the extracted identity embedding, we compared the cosine similarity of those from the same speaker (CSIS) with the cosine similarity of those from different speakers (CSID). The classification accuracy was computed for 2,000 speaker vectors, while the cosine similarity was calculated for 2,000 pairs of speaker vectors. 

As shown in Tab. \ref{tab1}, both the attribute embeddings extracted from real speaker vectors and those extracted from generated non-existent speaker vectors demonstrate high classification accuracy. Notably, since the SRVAE encoder is trained exclusively on real speaker vectors, these similar accuracies suggest that the non-existent speaker vectors generated by SRVAE reside in the same representation space as real speaker vectors. Furthermore, the CSIS and CSID values for the real speaker vectors show a gap of 0.653, demonstrating that the identity embeddings extracted by SRVAE can accurately represent speaker identity. Meanwhile, the gap for the generated speaker vectors is 0.633, a slight decrease, yet still sufficient to indicate that the proposed SRVAE is capable of extracting accurate identity embeddings from the modified non-existent speaker vectors.




\begin{table*}[htbp]
  \begin{center}
  \caption{Evaluation Results of Speech Gender Consistency, Identity Consistency, Intelligibility, and Quality}
  \label{tab2}
  \begin{tabular}{c|c c c c c c c} 
  \toprule [1pt]
   Model & SACC-gender $\uparrow$ & ICMOS $\uparrow$ & QCMOS $\uparrow$ & SIMS & SIMD & SIMS-age & SIMS-gender\\ 
  \midrule 
            GT           &88.5\%&+0.29&+0.12& 0.934 &0.596& n/a & n/a\\
        Baseline         &34.3\%&-0.06&-0.11& 0.913 &0.598&0.900&0.940\\
  \midrule 
        Proposed         &84.2\%& 0   &  0  & 0.766 &0.605&0.813&0.707\\
         -TSVC           &45.0\%&-0.27&-0.48& 0.807 &0.609&0.826&0.796\\
         -SRVAE          &57.5\%&-0.12&-0.21& 0.835 &0.613&0.851&0.800\\
        
  \bottomrule [1pt]
  \end{tabular}
  \end{center}
\vspace{-12pt}
\end{table*}
\vspace{-14pt}

\subsection{Evaluations of Generated Speech}
At the speech level, we conducted several subjective tests to assess the generated speech in terms of speech quality, attribute consistency, and identity consistency. For speech quality, we used Quality Comparative Mean Opinion Scores (QCMOS) to assess sound quality and naturalness, and Intelligibility Comparative Mean Opinion Scores (ICMOS) to evaluate pronunciation, as the Voxceleb2 dataset does not include text transcriptions. For gender consistency, we randomly assigned the generated and real samples for participants to predict the gender of the speaker. For age evaluation, we first asked participants to predict the speaker age of real speech without interval division. The most accurate age interval for human perception was then determined as 0-12, 12-25, 25-55, and $>$55, based on the prediction results. Finally, given the original speech and its age interval as a reference, participants were asked to predict the age interval of the modified speech. For identity consistency, participants were presented randomly with a modified or real speech sample from speaker A as a reference and were asked to select the speech with the same speaker identity from a test pair, which included speech from speaker B and another speech from speaker A. All subjective experiments involved at least 15 participants, with each being assigned 100 test questions. Additionally, we used the WavLM-based SR model to calculate speaker similarity between the speech from the same speaker (SIMS) and different speakers (SIMD), as in the speaker vector evaluation.

\textbf{Results on speech quality}: As shown in Tab. \ref{tab2}, the proposed model performs best in terms of intelligibility and quality of generated speech, but there is a 0.29 gap compared to real audio in ICMOS. We believe this is due to the SPPG model's lack of robustness when dealing with in-the-wild speech. In addition, the baseline model ranks second in this evaluation. We observe that it overfitted the training data, resulting in more than half of the modified speech being very similar to the source speech. Finally, the comparison with the ablative models demonstrates the effectiveness of our proposed approaches in modeling the natural expression of speaker attributes in speech.

\begin{figure}
\centerline{\includegraphics[width=0.95\linewidth]{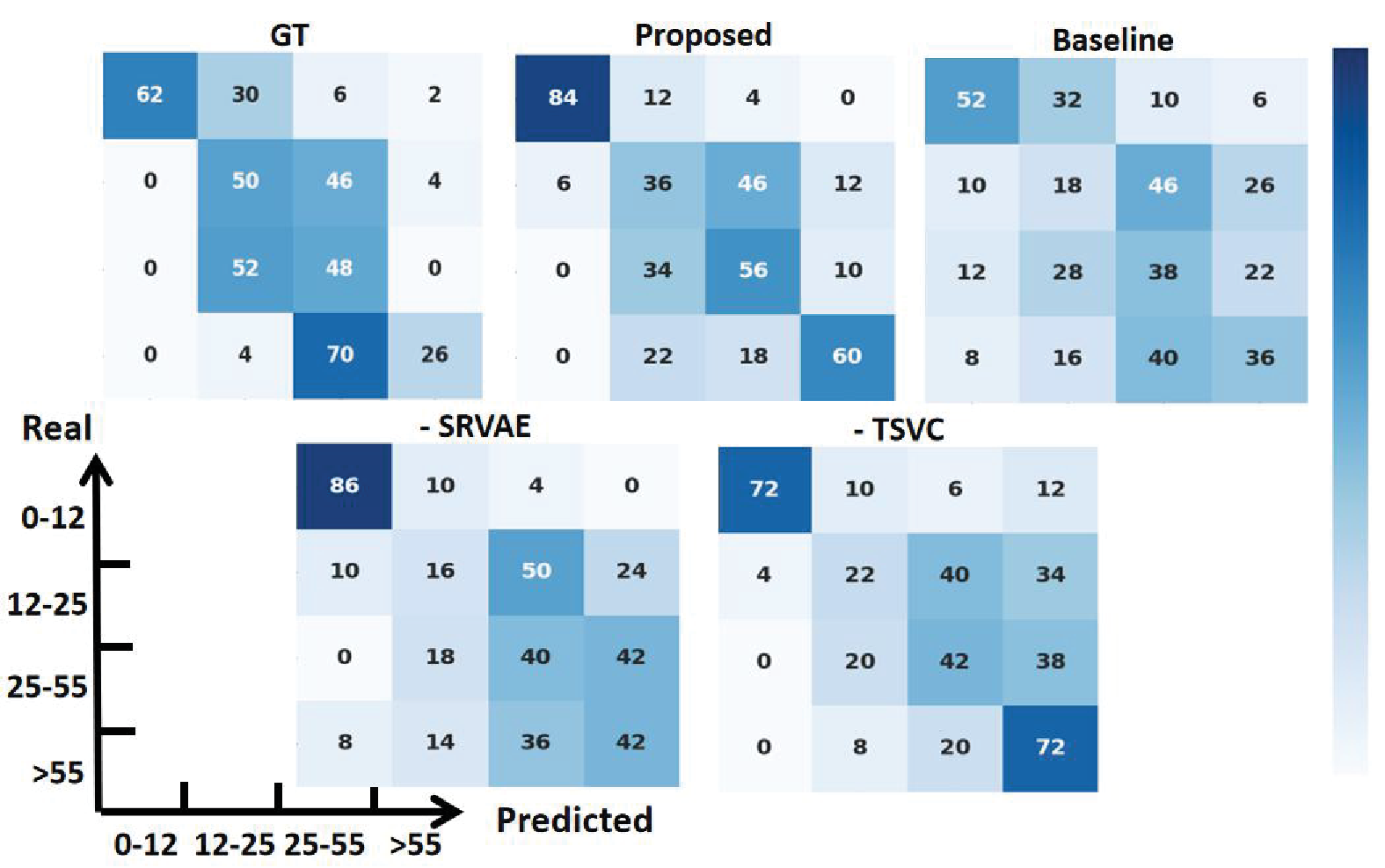}}
\caption{Confusion matrix for subjective prediction of speaker age.}
\label{fig2}
\end{figure}

\textbf{Results on attribute consistency}: 
Our method achieves a gender classification accuracy that is second only to ground truth, 49.9\% higher than the baseline model, as shown in Tab. \ref{tab2}. In age evaluation, our method obtains a confusion matrix most similar to the ground truth, while the confusion matrix of baseline exhibits a more diffuse distribution, as illustrated in Fig. \ref{fig2}. Additionally, after ablating SRVAE and TSVC, the predictive accuracy of gender decreases by 26.7\% and 39.0\%, respectively. Similarly, the consistency between the subjectively predicted age intervals and the predefined labels also decreases. These results further demonstrate the effectiveness of SRVAE and TSVC for attribute control.


\begin{figure}
\centerline{\includegraphics[width=1.0\linewidth]{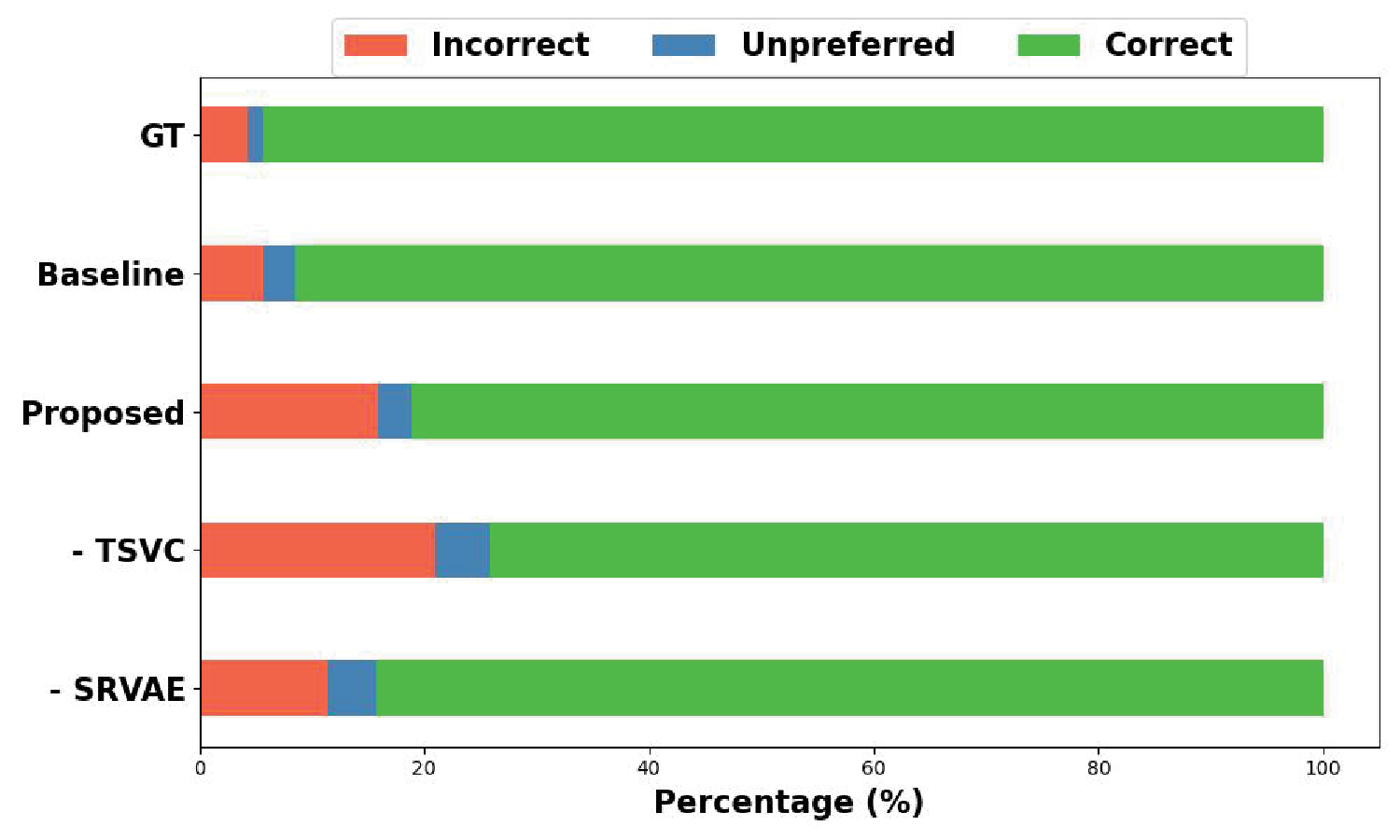}}
\caption{AB test results of speaker identity.}
\label{fig3}
\end{figure}

\textbf{Results on identity consistency}: Tab. \ref{tab2} shows the objective results using SR model. All methods achieve similar SIMD scores, while their SIMS exhibit a decreasing trend as attribute consistency increases. This trend is more pronounced when testing on speech with modified gender. This phenomenon can be explained by the fact that it is challenging for humans to recognize a speech pair with a large gap in fundamental frequency as belonging to the same identity. Although there is a decreasing trend in SIMS scores, they still exhibit a significant gap compared to SIMD, indicating that modified speech can still be recognized as originating from the source speaker by the SR model. Fig. \ref{fig3} shows the subjective results. The probability of participants selecting the correct audio is positively correlated with the gap between SIMS and SIMD in Tab. \ref{tab2}, except for "- TSVC". We believe this is because the poor quality of the speech generated by it affects the perception of participants.


\section{Conclusion}

In this letter, we introduce a two-step method for controlling speaker attributes in speech. Experimental results show that SRVAE can generate non-existent speaker vectors with predefined attributes that benefit from cyclic consistency training. By combining SRVAE with TSVC, we achieve speaker age and gender control at the speech level, while preserving as much of the original speaker's identity and speech quality as possible. This capability suggests that our method has potential applications in areas such as audiobooks, video dubbing, and privacy protection. This approach can also be applied to the text-to-speech task simply by replacing the input semantic features with text. Further work could focus on exploring whether the diverse speech generated by our method can improve tasks such as speaker recognition and self-supervised speech representation learning.

\bibliographystyle{IEEEbib}
\bibliography{ref}

\end{document}